\documentclass{PoS}
\usepackage{siunitx}
\usepackage{hepnames}

\title{Top quark mass measurements with the CMS experiment at the LHC}

\ShortTitle{Top quark mass measurements with CMS}

\author{\speaker{Simon Spannagel}%
  \thanks{on behalf of the CMS Collaboration.}
  \\
  Deutsches Elektronen-Synchrotron DESY\\
  E-mail: \email{simon.spannagel@desy.de}}

\abstract{Measurements of the top quark mass are presented, obtained from CMS data collected in proton proton collisions at the LHC at centre-of-mass energies of \SI{7}{TeV} and \SI{8}{TeV}.
  The mass of the top quark is measured using several methods and channels, including the reconstructed invariant mass distribution of the top quark as well as measurements based on charged particle information. The dependence of the mass measurement on the kinematic phase space is investigated. The results of the various channels are combined and compared to the world average. The top mass is extracted from the inclusive top quark pair production cross section measured at CMS.}

\FullConference{XXIV International Workshop on Deep-Inelastic Scattering and Related Subjects\\
         11-15 April, 2016\\
         DESY Hamburg, Germany}

\newcommand{\mt}{\ensuremath{m_{\Ptop}}\xspace}
\newcommand{\mtp}{\ensuremath{m_{\Ptop}^{\mathrm{pole}}}\xspace}

\newcommand{\ttbar}{\ensuremath{\Ptop\APtop}\xspace} % t-tbar
\newcommand{\pT}{\ensuremath{p_{T}}\xspace} % p_T
\newcommand{\units}[4]{\ensuremath{#1\pm#2\,\text{(stat)}\pm#3\,\text{(syst)}\,\text{#4}}}

\begin{document}

\section{Introduction}

The mass of the top quark \mt is a fundamental parameter of the standard model.
Together with the Higgs and \PW boson masses it provides a consistency test of the electroweak sector and has strong implications on the overall stability of the electroweak vacuum via the Higgs quartic coupling.
A deviation of top quark mass by a few GeV would predict an unstable vacuum and thus indicate the presence of physics beyond the Standard Model at the respective scale.

The mass of top quark has been measured by all major collider experiments at both the Tevatron and the LHC.
The measurements have been combined to a world average yielding a mass of $m_t = \units{173.34}{0.27}{0.71}{GeV}$~\cite{combination-tmass}.
The most recent combination of direct measurements by the CMS experiment~\cite{JINST-cms} as well as alternative methods are presented in the following.

\section{Direct Measurements of the Top Quark Mass}

Direct measurements obtain the top quark mass from a kinematic reconstruction of the decay products in \ttbar events.
These decays are typically divided into three different channels which are characterized by the decays of the \PW bosons following the decay $\Ptop \rightarrow \PW \Pbottom$.
The subsequent sections briefly present the top quark mass measurements from the three channels as well as their combination.
More details on the analyses are provided in~\cite{arXiv:1509.04044}.

\subsection{All-Jets}
The all-jets channel, with both \PW bosons decaying hadronically, features the largest branching ratio and allows for a full kinematic reconstruction as no neutrinos are produced.
However, the large number of jets involved leads to a large combinatorics in assigning the jets to particular partons and the large QCD multi-jet background requires tight selection criteria.

\begin{figure}[bp]
  \centering
  \includegraphics[width=.45\textwidth]{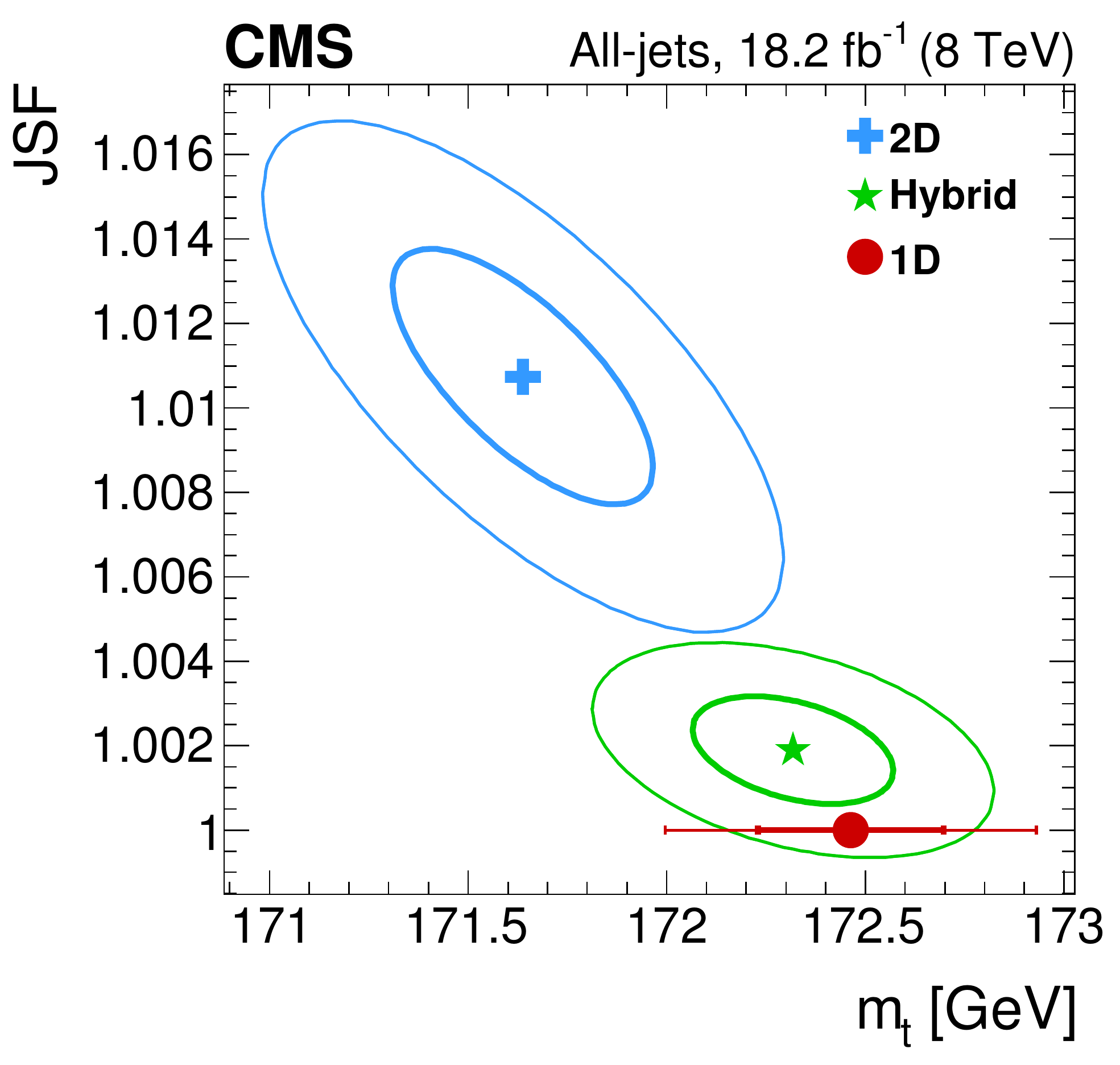}%
  \includegraphics[width=.45\textwidth]{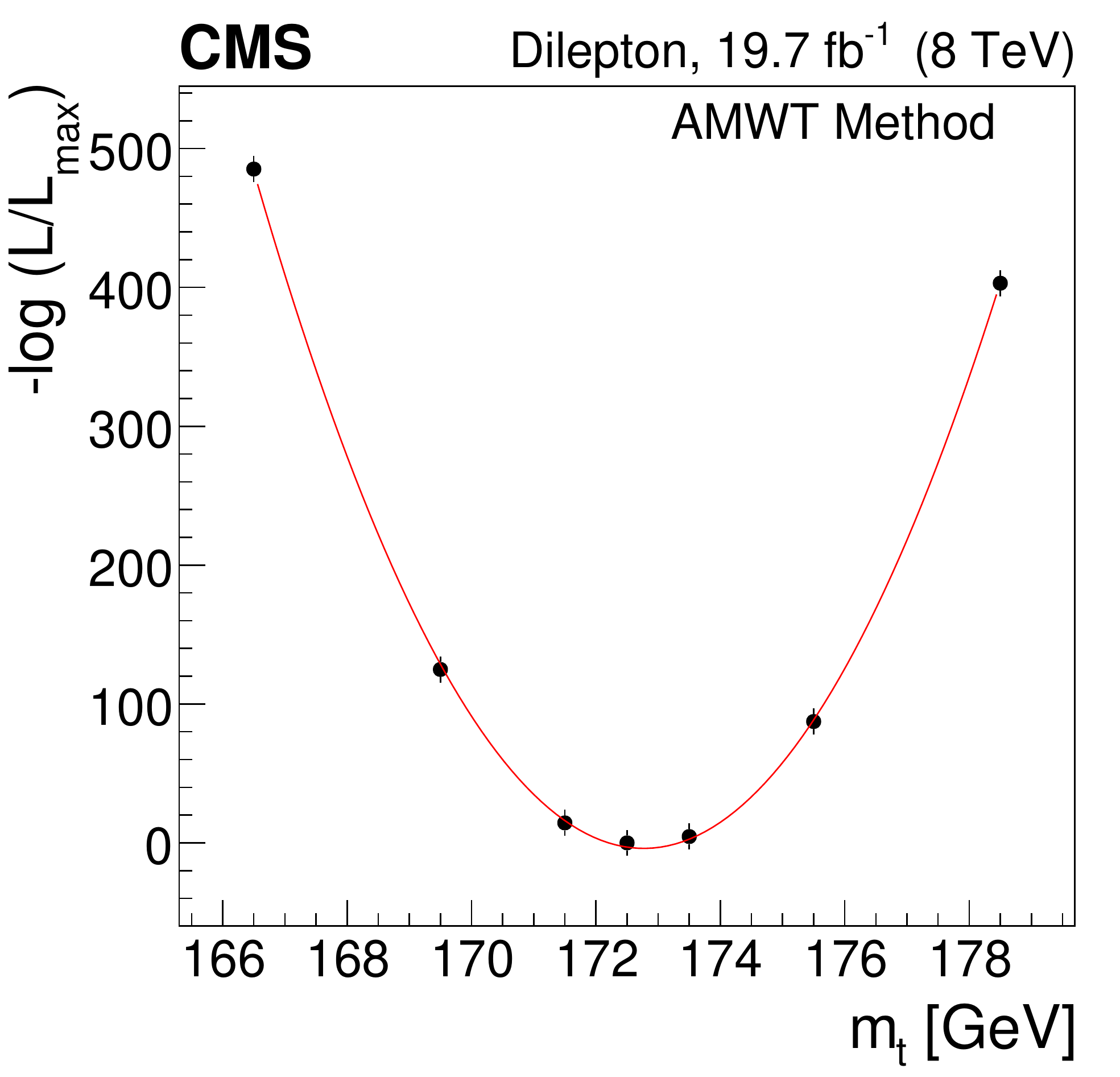}
  \caption{Top quark masses as measured using the ideogram method in the all-jets channel~(left). Likelihood fit of the AMWT method used for the dileptonic channel~(right)~\cite{arXiv:1509.04044}.}
  \label{fig:direct}
\end{figure}

The top quark mass is measured using the ideogram method.
A per-event likelihood is calculated taking into account the expected contributions from correct/wrong matching of the jets to partons as well as from backgrounds.
A kinematic reconstruction provides different solutions for the jet-parton assignment and the solution which best fits the \ttbar hypothesis is chosen.
The likelihood values are combined in order to perform the measurement in one or two dimensions, or following a hybrid approach.
In the 1D method, \mt is directly inferred from the templates, while the 2D method allows to determine \mt and a global jet scale factor (JSF) simultaneously.
While the latter does not use any prior for the JSF, the hybrid method combines the two options by imposing a Gaussian constraint on the JSF with a width corresponding to the respective jet calibration uncertainty.

Figure~\ref{fig:direct}~(left) shows the three measured values for \mt with their statistical and total uncertainties.
The hybrid method yields a mass of $\mt^{hyb} = 172.32\pm0.25\,\textrm{(stat+JSF)}\,\pm0.59\,\textrm{(syst)\,GeV}$.

\subsection{Lepton+Jets}

The lepton+jets channel, where one of the \PW bosons is required to decay leptonically, provides a modest branching ratio while allowing for more relaxed selection criteria.
The event is identified via the top quark associated to the leptonically decaying \PW boson, while the mass is measured from the fully reconstructed hadronic top quark candidate.
Since selected events usually feature four jets or more, the combinatorics remain an issue.

The mass is measured using the three different ideogram methods described in the previous section.
The impact of events with wrongly assigned jets is mitigated by using the goodness-of-fit probability $P_{\textrm{gof}}$ from the kinematic reconstruction as an additional event weight.
The hybrid measurement yields $\mt^{hyb} = 172.35\pm 0.16\,\textrm{(stat+JSF)}\,\pm 0.48\,\textrm{(syst)\,GeV}$ with a total precision of \SI{0.3}{\percent} (\SI{0.51}{GeV}).
The dominant systematic uncertainty is the jet energy correction uncertainty arising from the b quark fragmentation.

\subsection{Dileptonic}

The dileptonic channel is characterized by both \PW bosons decaying leptonically.
This channel features a very low fraction of background events of typically only a few percent and allows for a simplified combinatorics with only two lepton/b-jet association permutations.
However, a full event reconstruction is not possible due to the two neutrinos escaping detection, which can only be inferred via the missing energy in the transverse plane.

For this reason, the analytical matrix weighting technique (AMWT) is used to extract the top quark mass.
A scan of \mt is performed from 100 to \SI{400}{GeV} and the probability of observing a charged lepton with energy~$E$ in the rest frame of a top quark with the respective mass \mt is calculated.
Event weights are assigned using the determined probabilities and the observable $\mt^{\textrm{AMWT}}$ is chosen as the mass with the highest average sum of weights.

The top quark mass is measured by comparing the observable to simulations with different \mt hypotheses.
The mass is determined from the minimum of the parabola fit to the individual binned likelihoods as shown in Figure~\ref{fig:direct}~(right).
A mass of $\mt = 172.82 \pm 0.19\,\textrm{(stat)}\,\pm 1.22\,\textrm{(syst)\,GeV}$ is measured from dileptonic \ttbar events using the AMWT method.

\subsection{Combination}

The above measurements have been combined with results from 2010, 2011 and 2012 using the BLUE method~\cite{blue} which takes correlations between the different measurements into account.
%The correlations are weighted by their uncorrelated uncertainties.
The combined top quark mass is $\mt = 172.44\pm 0.13\,\textrm{(stat+JSF)} \pm 0.47\,\textrm{(syst)\,GeV}$ and represents the currently most precise measurement of the top quark mass with a total precision smaller than~\SI{0.3}{\percent}.

%\begin{figure}
%  \centering
%  \includegraphics[width=.6\textwidth]{topmass_cms.pdf}
%\end{figure}

\section{Differential Measurements of the Top Quark Mass}

The large number of \ttbar events recorded by the CMS experiment allow to measure \mt differentially as a function of different kinematic variables~\cite{arXiv:1509.04044}.
These measurements provide insight into potential limitations of current event generators and facilitate searches for possible biases.
In order to obtain \mt differentially, the hybrid ideogram method described above is applied to a subset of the selected events, which are binned by the respective observable.

\begin{figure}[tbp]
  \centering
  \includegraphics[width=.45\textwidth]{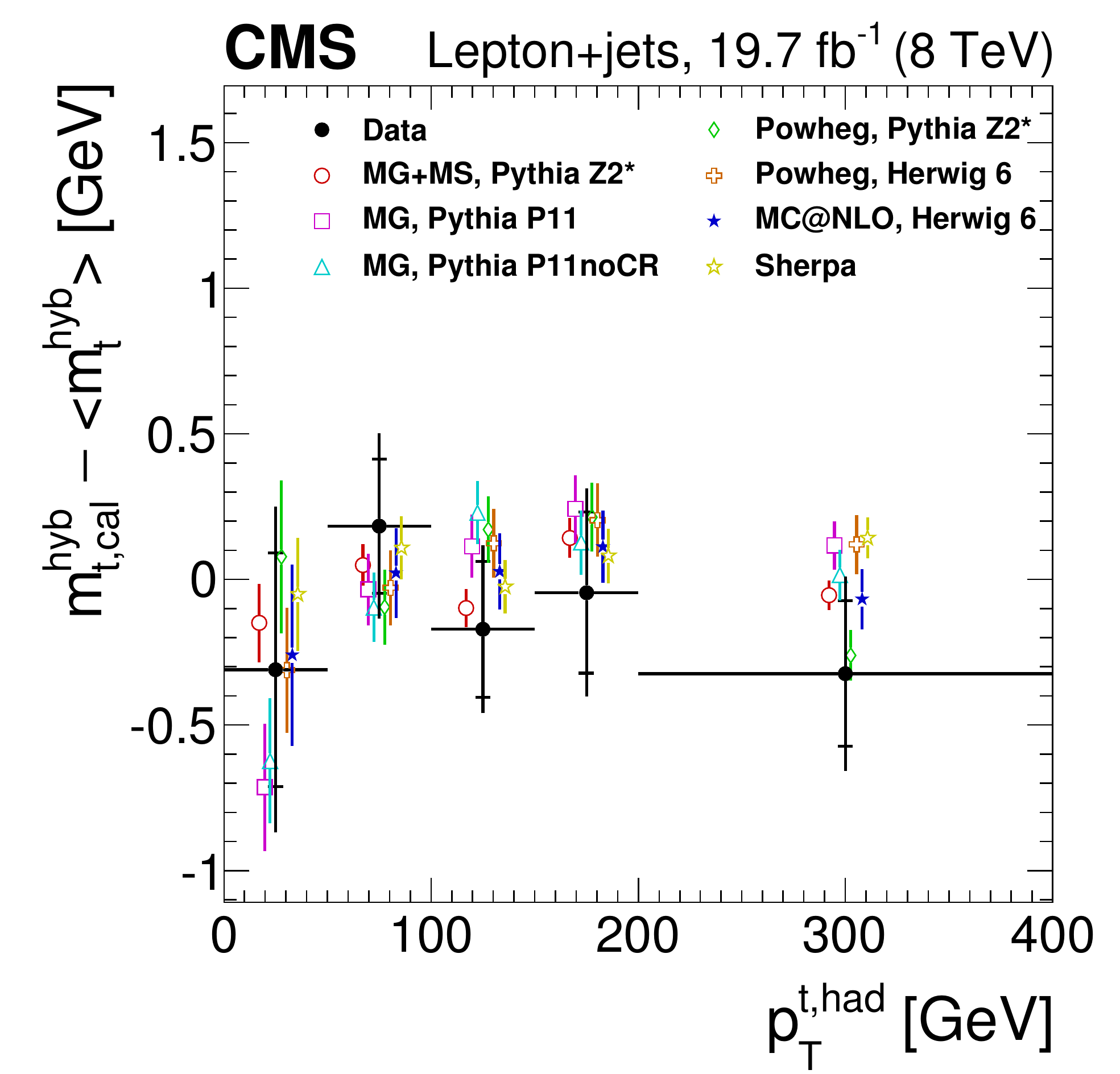}%
  \includegraphics[width=.45\textwidth]{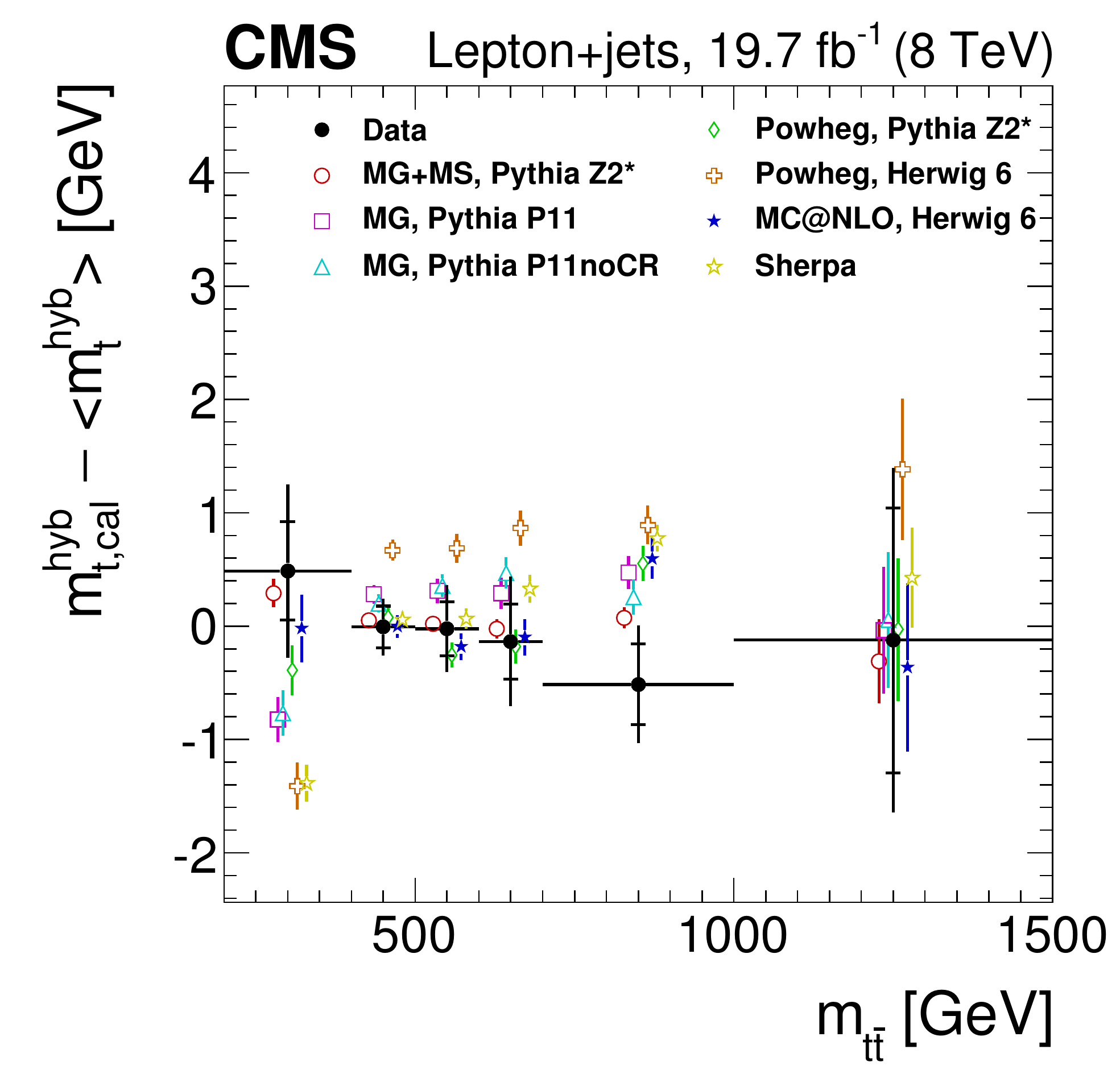}
  \caption{Differential measurements of the top quark mass as a function of the transverse momentum of the hadronic top quark~(left) and the invariant mass of the \ttbar system~(right)~\cite{arXiv:1509.04044}.}
  \label{fig:diff}
\end{figure}

Figure~\ref{fig:diff} shows two examples of differential measurements performed in the lepton+jets channel.
The left plot compares the measured top quark mass as a function of the transverse momentum of the hadronic top quark $\pT^{\Ptop, \mathrm{had}}$ with several predictions obtained from simulation, allowing to compare the description of the top quark \pT in the different generator programs with data.
Figure~\ref{fig:diff}~(right) shows a comparison of the measured and simulated top quark mass as a function of the invariant mass of the \ttbar system $m_{\ttbar}$, which facilitates testing the scale of the process.

\section{The Top Quark Mass from Charged Particles}

An interesting alternative to the direct top quark mass measurements is the determination of the mass solely from charged particles.
These methods aim at reducing the dependence of the measurement on detector calibrations such as jet energy corrections.
This is achieved by evaluating the information obtained from particle tracks only, while not reconstructing the actual top quark.

\begin{figure}[tbp]
  \centering
  \includegraphics[width=.45\textwidth]{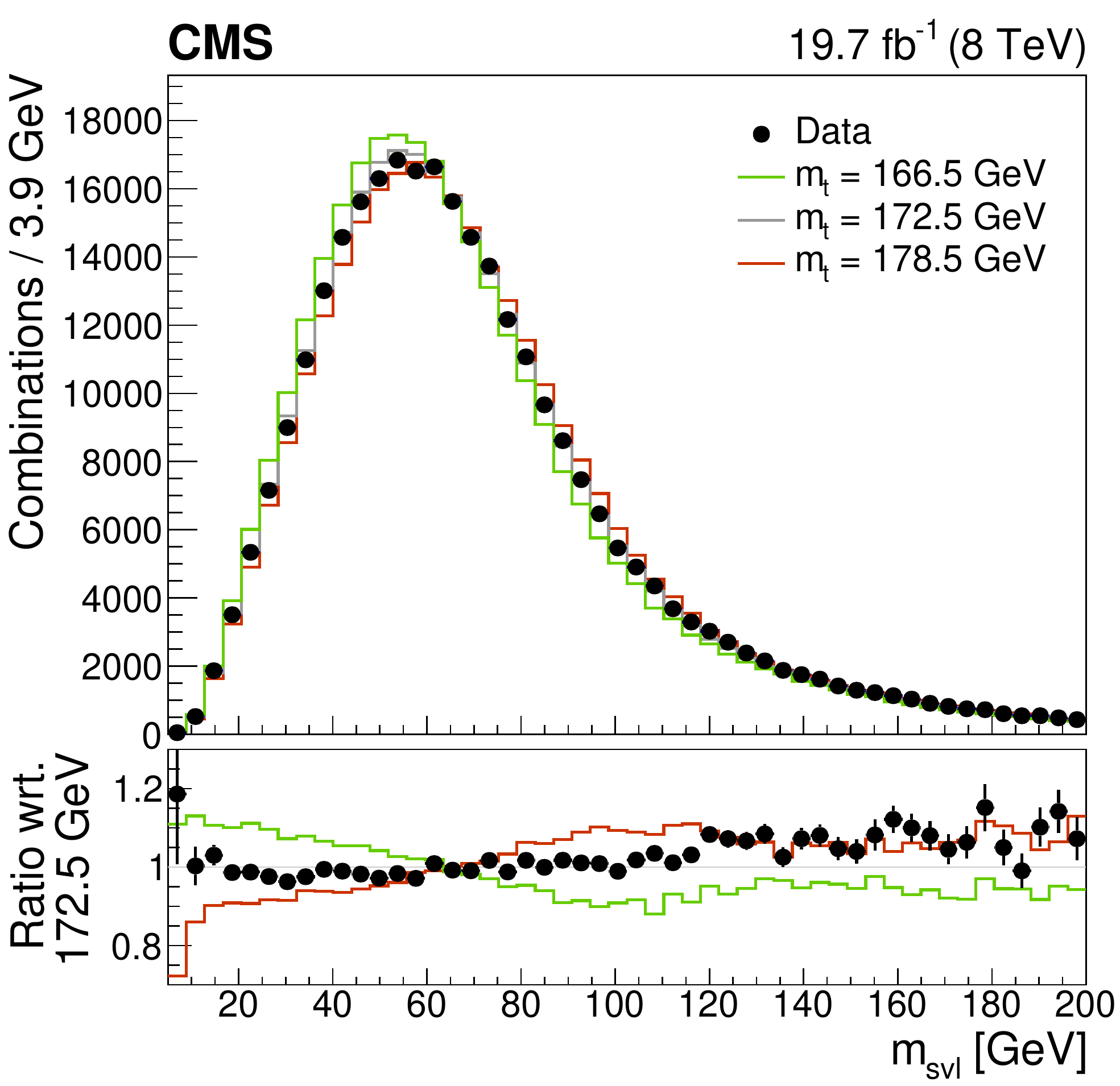}%
  \includegraphics[width=.45\textwidth]{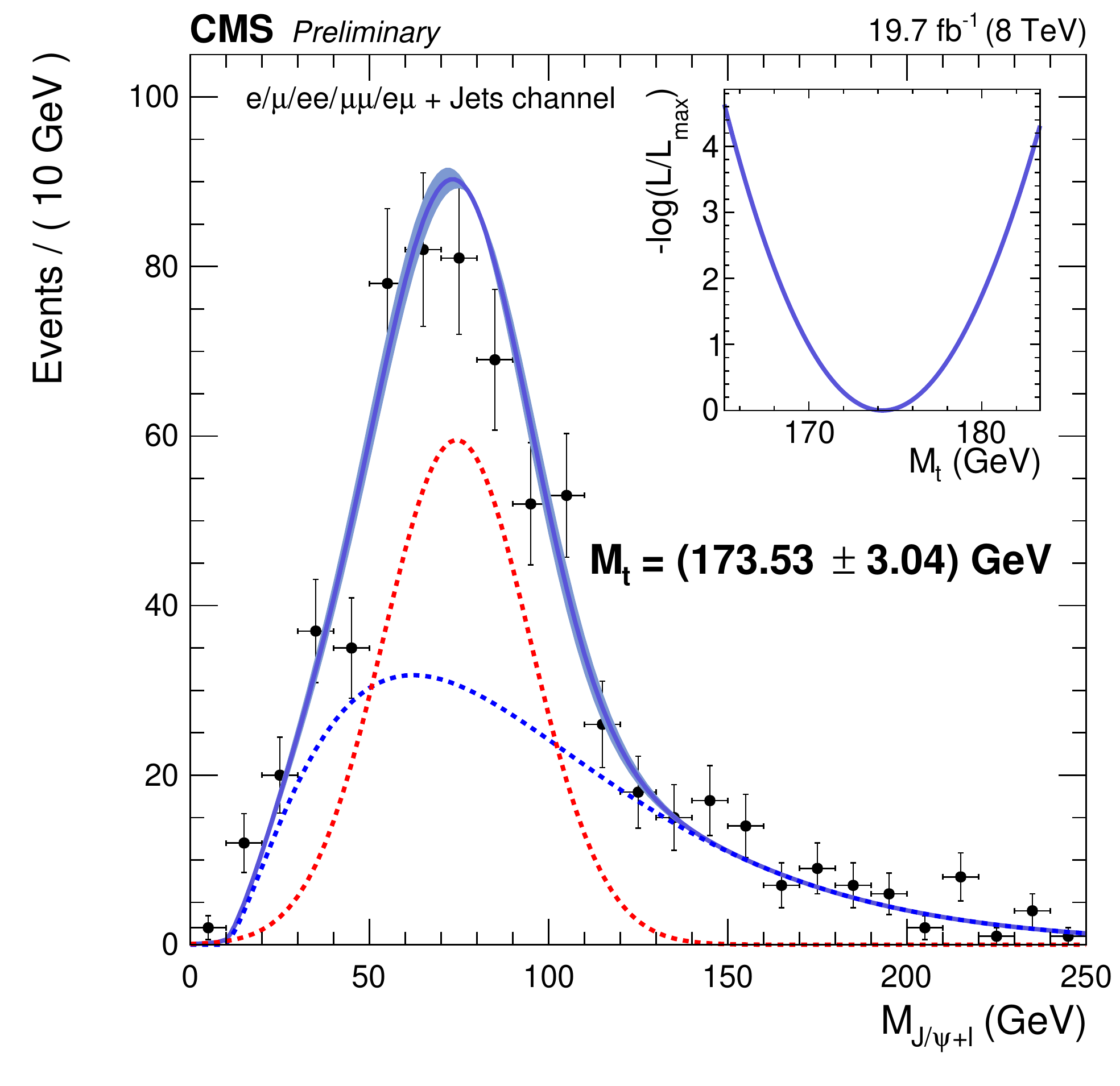}
  \caption{Comparison of the $m_{svl}$ distribution with simulations with different top quark mass values~(left)~\cite{arxiv:1603.06536}. Fit of the lepton+\PJpsi invariant mass with signal and background contributions~(right)~\cite{CMS-PAS-TOP-15-014}.}
  \label{fig:charged}
\end{figure}

One observable considered is the invariant mass of the lepton from the \PW decay and the secondary decay vertex of the b quark, $m_{svl}$, as shown in Figure~\ref{fig:charged}~(left)~\cite{arxiv:1603.06536}.
Here, events from the dileptonic and lepton+jets \ttbar channels are combined and all combinations of secondary vertex and lepton assignments are taken into account.
The observed $m_{svl}$ distribution is fitted %including all possible components of correct and wrong jet-parton matching as well as background contributions.
and the top quark mass is determined from the maximum likelihood of all channels considered.
The measured top quark mass amounts to $\mt = 173.68 \pm 0.20\,\textrm{(stat)}^{+1.58}_{-0.97}\,\textrm{(syst)\,GeV}$ with an uncertainty arising from experimental sources of only \SI{0.44}{GeV}.

Another observable is the invariant mass of the lepton and a $\PJpsi$ meson, $m_{\PJpsi+l}$, as shown in Figure~\ref{fig:charged}~(right)~\cite{CMS-PAS-TOP-15-014}.
While this measurement is still statistically limited due to the small branching fraction of about \SI{0.3}{\percent}, it features small systematic uncertainties and is a promising candidate for future measurements with larger statistics.

\section{Indirect Measurement from the Inclusive Cross Sections}

Measuring the top quark mass from the mass dependence of the inclusive \ttbar production cross section provides access to the top quark pole mass and does not rely on a correct kinematic reconstruction of the \ttbar system from decay products.
The inclusive production cross section can be calculated at NNLO in QCD and comparing the prediction at a given center-of-mass energy to the corresponding measurement allows to directly measure the pole mass.

\begin{figure}[tbp]
  \centering
  \includegraphics[width=.55\textwidth]{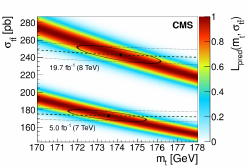}%
  \caption{Measurement of the top quark pole mass from the inclusive \ttbar production cross section using calculations at NNLO in QCD at $\sqrt{s} = \SI{7}{TeV}$ and \SI{8}{TeV}~\cite{arxiv:1603.02303}.}
  \label{fig:xsec}
\end{figure}

Figure~\ref{fig:xsec} presents the latest combined  measurement of the pole mass from the inclusive \ttbar cross section %and an value of $\alpha_{s} = 0.118 \pm 0.001$.
%The combined measurement
at $\sqrt{s} = \SI{7}{TeV}$ and \SI{8}{TeV} using the NNPDF3.0~\cite{nnpdf3} parton distribution functions.
The measurement yields $\mtp = 173.8 ^{+1.7}_{-1.8}\,\textrm{GeV}$~\cite{arxiv:1603.02303}.

\section{Summary and Outlook}

The top quark mass is an important parameter to the standard model and the large \ttbar production cross section at the LHC allows precision measurements of this property.
The latest direct measurement of the top quark mass performed by the CMS experiment provides \mt with a precision of \SI{0.3}{\percent}.
The dominating uncertainties are jet energy corrections arising from the detector calibration and modeling uncertainties of the underlying theory.

Differential mass measurements provide additional insight into the process and allow detailed comparisons with various Monte Carlo generators.
New methods such as the measurement of the mass from charged particles allow to reduce the persistent uncertainties arising from experimental sources such as jet energy calibrations.
The indirect measurement from the inclusive \ttbar production cross section allows to measure the top quark pole mass in a well-defined renormalization scheme.

Alternative methods such as measurement from differential cross sections are investigated and represent interesting candidates for the data collected at the increased LHC center-of-mass energy of $\sqrt{s} = \SI{13}{TeV}$.

\bibliographystyle{unsrt}
\bibliography{bibliography}

\end{document}